\documentclass[aps,twocolumn,nofootinbib,floatfix]{revtex4} 
\input epsf
\newcommand{\sfig}[2]{
\centerline{ \epsfxsize = #2 \epsfbox{#1} }
		}
\newcommand{\Sfig}[2]{
	\begin{figure}[thp]
	\sfig{#1.eps}{0.95\columnwidth}
	\caption{{\small #2}}
	\label{fig:#1}
	\end{figure}
}

\def\cmm2{{\,\rm cm^{-2}}}
\def\cm2{{\,{\rm cm}^2}}
\def\cmm3{{\,{\rm cm}^{-3}}}
\def\gcmm3{{\,{\rm g\,cm^{-3}}}}

\def\fun#1#2{\lower3.6pt\vbox{\baselineskip0pt\lineskip.9pt
  \ialign{$\mathsurround=0pt#1\hfil##\hfil$\crcr#2\crcr\sim\crcr}}}

\def\ie{{\it i.e.} }

\def\be{\begin{equation}}
\def\ee{\end{equation}}
\def\bea{\begin{eqnarray}}
\def\eea{\end{eqnarray}}
\newcommand{\vs}{\nonumber\\}

\def\bfH{{\bf H}}
\def\bfI{{\bf I}}

\def\bfe{{\bf e}}

\def\bfk{{\bf k}}
\def\bfr{{\bf r}}
\def\bfx{{\bf x}}
\def\hatbfn{{\hat{\bf n}}}

\def\ccdot{{\hskip-0.7pt\cdot\hskip-0.7pt}}
\def\ctimes{{\hskip-0.7pt\times\hskip-0.7pt}}
\def\cminus{{\hskip-0.7pt-\hskip-0.7pt}}

\newcommand{\eql}[1]{\label{eq:#1}}

\begin{document}

\title{Primordial Gravity Waves and Weak Lensing} 

\author{Scott Dodelson$^{1,2}$, Eduardo Rozo$^{3}$, and Albert Stebbins$^{1}$
}

\affiliation{$^1$NASA/Fermilab Astrophysics Center
Fermi National Accelerator Laboratory, Batavia, IL~~60510-0500}
\affiliation{$^2$Department of Astronomy \& Astrophysics, The University of Chicago, 
Chicago, IL~~60637-1433}
\affiliation{$^3$Department of Physics, The University of Chicago, 
Chicago, IL~~60637-1433}

\date{\today}

\begin{abstract}
Inflation produces a primordial spectrum of gravity waves in addition to the
density perturbations which seed structure formation. We compute the signature
of these gravity waves in the large scale shear field.  In particular, the
shear can be divided into a gradient mode (G or E) and a curl mode (C or
B). The
former is produced by both density perturbations and gravity waves, while the
latter is produced only by gravity waves, so the observations of a non-zero
curl mode could be seen as evidence for inflation.  We find that the expected
signal from inflation is small, peaking on the largest scales at
$l(l+1)C_l/2\pi < 10^{-11}$ at $l=2$ and falling rapidly there after. Even
for an all-sky deep survey, this signal would be below noise at
all multipoles.  Part of the reason for the smallness of the signal is
a cancellation on large scales of the standard line-of-sight effect and
the effect of ``metric shear.''

\end{abstract}
\maketitle


The theory of inflation was proposed over twenty years ago\,\cite{guth}. For the
first years after its discovery/invention, cosmologists worked out some of its
cosmological implications and particle physics implementations.  While this
work is still going on, the most exciting development in the last several years
has been the confirmation of some of inflation's basic predictions.  It now
appears that the universe is flat, the most robust prediction of
inflation\,\cite{flat}.  Further, observations of large scale structure and
anisotropies in the cosmic microwave background (CMB) strongly support the idea
that small, adiabatic, nearly scale-invariant fluctuations were present in the
early universe. Inflation predicts the existence of precisely this class of
perturbations\,\cite{Liddle}.

The case for inflation is not airtight though. Inflation explains why the
universe appears flat today even if the curvature is not exactly equal to zero.
But it is possible that the curvature really is exactly zero, so that inflation
is not needed to account for the apparent flatness. Similarly, the adiabatic,
scale-free perturbation spectrum predicted by inflation might have been laid
down by some other mechanism. Indeed, it is somewhat of an embarrassment for
proponents of inflation that this spectrum is called the
``Harrison-Zel'dovich-Peebles'' spectrum, named after the three eminent
cosmologists who first proposed it, long before inflation had been suggested.

Are there any signatures unique to inflation? Just as the primordial density
perturbations were produced by quantum fluctuations during inflation,
primordial gravity waves were produced by quantum fluctuations of the metric
during the inflationary epoch\,\cite{gravwave}.  If we were to detect these
gravity waves, the case for inflation would be strengthened considerably. After
the detection of anisotropies in the CMB by the COBE satellite\,\cite{cobe},
there was hope that the gravity waves would eventually be extracted from an
accurate measurement of the full anisotropy spectrum\,\cite{steinhardt}.  Recent
work has shown though that other cosmological parameters, especially late
reionization, can mimic the effects of gravity waves in the anisotropy
spectrum\,\cite{daniel}, making detection more difficult.

In 1996, two groups\,\cite{albert,uros} showed that gravity waves produce a
polarization pattern in the CMB that cannot be caused by scalar (density)
perturbations to first order in the amplitude.  The polarization field can be
decomposed into two modes.  One is technically a {\it scalar} pattern on the
sky, just as a gradient is, and the names given to such modes have variously
been ``scalar'', ``G'' (for gradient) , or ``E'' (for electric since an electric
field is a gradient of a scalar potential).  Another orthogonal set of modes
are {\it pseudoscalar} patterns on the sky, and can be obtained by rotating the
polarization at each direction in a {\it scalar} pattern by $45^\circ$.  Such
modes change sign under parity like a curl does, and names given to them have
variously been ``pseudoscalar'', ``C'' (for curl), and ``B'' (for magnetic
since a magnetic field is also odd under parity).  The polarization pattern
produced by a single sinusoidal\footnote{or pseudo-sinusoidal in the case of an
FRW cosmology with non-zero curvature} density perturbation must be
rotationally symmetric about the wavenumber $\bfk$ and also symmetric under
reflections in the plane perpendicular to $\bfk$.  Such a pattern cannot
contain any C ($\equiv$B) modes.  In linear theory the polarization pattern
produced by density perturbations is just the sum of the pattern from
individual sine waves and hence we derive the general rule that scalar
perturbations produce no C modes\footnote{Nonlinearities can lead to C modes
even from scalar inhomogeneities but will be most important on small angular
scales.  The gravity wave signal discussed here is on large
angular scales.}.  This symmetry argument does not apply to gravity waves which
do not have the same reflection symmetry because each gravity wave is polarized
in a particular direction.  Gravity waves do produce C mode polarization
patterns in the CMB. Thus detection of C modes can provide fairly unambiguous
evidence for gravity waves, thereby verifying a unique prediction of inflation.
It will take quite a while to reach the sensitivity needed to make this test a
powerful probe of inflation\,\cite{kinney}.  Along the way, there are many
systematic effects which might prove to be spoilers, prominent among them the
possibility that foregrounds are polarized.

The pattern of shearing of images due to gravitational, or any other kind of,
lensing is described by a symmetric traceless tensor field on the sky, just as is the
polarization pattern, and like polarization can be decomposed into G and
C modes.  The gravitational field causing the lensing can be scalar (caused by
density inhomogeneities), vector (caused by vorticity), or tensor (gravity
waves).  The symmetry argument stated above tells us that C mode shear pattern
can be produced by gravity waves and not by density
inhomogeneities\,\cite{stebbins96}.  As with CMB polarization, we expect the
dominant shear to be G modes produced by density inhomogeneities, but we can
hope to detect gravity waves through C modes in the shear pattern.  This raises
the question of whether the primordial gravity waves produced by inflation
might be detectable by measuring the C mode of shear.  Here, we explore this
question. 

One measures the shear field by correlating the shapes of distant galaxies, in
particular the ellipticity of galactic light distribution on the sky\,\cite{wl},
and as with polarization we can expect various other effects to mimic C mode
shear from gravity waves in such a measurement.  We will not address the issue
of ``foregrounds'' in this paper, but as we will show (see also 
Ref.~\cite{KaiserJaffe97}) the amplitude of shear produced by gravity
waves is so small that the expected signal is below even the statistical
noise expected in the largest experiment imaginable.


Gravitational lensing is caused by the deflection of light trajectories due to
metric perturbations.  The metric of a flat Friedman-Robertson-Walker cosmology
in synchronous coordinates\footnote{Linearized tensor (gravity wave)
perturbations always lead to synchronous coordinates, but other coordinate
choices (gauges) are available for vector (vorticity) and scalar (density)
perturbations.} is given by
\be
g_{\mu\nu} = a^2\left( \matrix{-1 & 0 \cr
				0 & \bfI+\bfH } \right)
\ee
where $a(\eta)$ is the scale factor of the universe; $\eta$ is the conformal
time variable; and the $3\times3$ matrices $\bfI$ and $\bfH$ are respectively 
the identity matrix and the metric perturbation which is symmetric
$\bfH=\bfH^{\rm T}$. The deflection of light is described by the geodesic
equation,
\be
\ddot{\bfr}=
\frac{1}{2} \left( \dot{\bfr} \cdot \dot{\bfH} \cdot \dot{\bfr}\right) \dot{\bfr} 
           -(\bfI+\bfH)^{-1}\cdot\left(
                  \dot{\bfr}\cdot{d\over d\eta}\bfH
                  -{1\over2}\nabla(\dot{\bfr}\cdot\bfH\cdot\dot{\bfr})\right)
\ee
where $\dot{}=\frac{\partial}{\partial\eta}$ and
${d\over d\eta}=\frac{\partial}{\partial\eta}+\dot{\bfr}\cdot\nabla$.  A full
solution of this equation (ray-tracing) is unnecessary since we may not
only linearize the equation in  $\bfH$, but also linearize the solution in
$\bfH$ by evaluating the gravitational accelerations on an unperturbed
trajectory, \ie the Born approximation.  A ``Born trajectory'' arriving at the
origin at $\eta=\eta_0$ coming from direction $\hatbfn$ is
\bea
&&\hskip-10pt
\bfr(\eta)=\left(\hatbfn-{1\over2}\bfH_0\cdot\hatbfn\right)\,(\eta_0-\eta)
-\int_\eta^{\eta_0} d\eta'\times\vs
&&\hskip-10pt
\left[\bfH\cdot\hatbfn+{1\over2}(\eta'-\eta)\,
      \left((\hatbfn\cdot\dot{\bfH}\cdot\hatbfn)\,\hatbfn
             -\nabla(\hatbfn\cdot\bfH\cdot\hatbfn)\right)
      \right]_{(\eta',\bfx')}
\eea
where $\bfx'=(\eta_0-\eta')\hatbfn$.  The initial conditions are chosen such
that the trajectory is light-like to the angle between two trajectories
$\hatbfn$ and $\hatbfn'$ intersecting at the origin is
$\cos^{-1}\hatbfn\cdot\hatbfn'$ (both to 1st order in $\bfH$).

One may decompose the perturbation of the trajectory into a displacement along
the line-of-sight (LOS), \ie a time delay; and a displacement perpendicular to
the LOS, $\delta\bfr_\perp=\bfr-(\hatbfn\cdot\bfr)\,\hatbfn$, which describes
all lensing effects.  It is convenient to convert from displacement in
coordinate distance to displacement in angle on the sky by defining
$\vec{\Delta}=\delta_\perp{\bfr}/(\eta_0-\eta)$, which can be thought of as
a 2-d vector in the tangent space of the direction sphere parameterized by
$\hatbfn$.

Following \cite{stebbins96} the displacement pattern on the direction sphere
described by $\vec{\Delta}$ can alternately be described in terms of the
convergence, $\kappa=-{1\over2}\Delta^{a}{}_{:a}$, which is the covariant
divergence of $\vec{\Delta}$; and the rotation
$\omega={1\over2}(\Delta_a\epsilon^{ab})_{:b}$, which is the is the covariant 
divergence of $\vec{\Delta}$ rotated by $90^\circ$ .  The G and C modes are
given by $\kappa$ and $\omega$, respectively.  In constrast to the G modes,
where the gravity wave contribution will be dwarfed by the density
perturbation, the C modes will have no contribution from density perturbations
on large scales, so we are interested in $\omega$ and not $\kappa$.
Rexpressing the 2-d relation $\omega={1\over2}(\Delta_a\epsilon^{ab})_{:b}$
in terms of the 3-d trajectory we find\footnote{One can see from this equation
that a scalar (density) perturbation does not contribute to $\omega$ and
hence to C modes.}
\bea\label{BornOmega}
\omega(\hatbfn,\eta)
&=&-{1\over2}{1\over\eta_0-\eta}
             \hatbfn\cdot\left(\nabla_\hatbfn\times\bfr(\hatbfn,\eta)\right)\cr
&=&{1\over2}\int_\eta^{\eta_0}d\eta'
                \left[\hatbfn\cdot(\nabla\times\bfH)\cdot\hatbfn
                       \right]_{\left(\eta',\hatbfn(\eta_0-\eta')\right)}
\eea
using the Born solution.  Henceforth one can consider $\eta$ as being a measure
of the distance to the background galaxies whose shear we measure.

	The contribution of the gravity waves (tensor modes) to $\bfH$ is
transverse and traceless, \ie Tr\,$\bfH=0$, $\nabla\cdot\bfH=0$, and one can
Fourier decompose the tensor constribution as
\be\label{FourierDecomposition}
\bfH^{\rm (T)}(\bfx,\eta)={1\over(2\pi)^{3\over2}}\int d^3\bfk\,
e^{i\bfk\cdot\bfx}\sum_{i=1}^2\tilde{\bfH}^{\rm (T)}_{(i)}(\bfk,\eta)
\ee
where
\be\label{FourierAmplitude}
\tilde{\bfH}^{\rm (T)}_{(i)}(\bfk,\eta)=
T_{\rm(T)}(k,\eta)\,\tilde{H}_{(i)}(\bfk)\bfe_{(i)}(\bfk)\ ,
\ee
$k=|\bfk|$, we define the transfer function such that $T_{\rm(T)}(k,0)=1$, and
$i$ sums over the two (linear) polarization states defined by the polarization
tensors, $\bfe_{(i)}$, which obey
${\rm Tr}\,\bfe_{(i)}={\rm Im}\,\bfe_{(i)}=0$, $\bfe_{(i)}=\bfe^{\rm T}_{(i)}$,
$\bfk\cdot\bfe_{(i)}(\bfk)=\bf0$, 
${\rm Tr}\left(\bfe_{(i)}(\bfk)\cdot\bfe_{(j)}(\bfk)\right)=2\delta_{ij}$.
Assuming isotropy and no preferred handedness we may define a power spectrum:
\be
\left\langle\tilde{H}_{(i)}(\bfk)\tilde{H}_{(j)}^*(\bfk')\right\rangle
=(2\pi)^3P_{\rm(T)}(k)\,\delta_{ij}\,\delta^{(3)}(\bfk-\bfk')\ .
\ee
In inflationary models with Hubble parameter $H_{\rm I}$ during inflation
\be
P_{\rm(T)}(k)={8\pi\over(2\pi)^3}
              \left({H_{\rm I}\over M_{\rm Planck}}\right)^2 k^{-3} \ .
\ee
which gives equal metric perturbation on all scales. Presently CMB
observations limit\,\cite{melchiorri} $H_{\rm I}\le2\times 10^{14}$GeV.

	An integral required to compute $\omega$ is
\bea
\Omega_{(i)}(\bfk,\hatbfn,\eta)&=&-{i\over2}k\sin^2\theta\,\sin2\phi\cr
&&\times\int_\eta^{\eta_0}d\eta'e^{ik\,\cos\theta\,(\eta_0-\eta')}T(k,\eta)
\eea
where the two angles, $\theta$ and $\phi$, are defined by
$\hatbfn\cdot\bfk=k\,\cos\theta$ and 
$\hatbfn\cdot\left(\bfk\times\bfe_{(i)}(\bfk)\right)\cdot\hatbfn
=k\,\sin^2\theta\,\sin2\phi$.
Eq.s~(\ref{BornOmega},\ref{FourierDecomposition},\ref{FourierAmplitude})
becomes
\be
\omega(\hatbfn,\eta)
={1\over(2\pi)^{3\over2}}\int d^3\bfk\,
\sum_{i=1}^2\tilde{H}_{(i)}(\bfk)\,\Omega_{(i)}(\bfk,\hatbfn,\eta)
\ee
If we define
\be
\tilde{\omega}_{(l,m)}(\eta)
=\int d^2\hatbfn\,Y_{(l,m)}^*(\hatbfn) \omega(\hatbfn,\eta)
\ee
then
\bea
C_l^\otimes(\eta)&=&{1\over2l+1}\sum_{m=-l}^l 
\left\langle|\tilde{\omega}_{(l,m)}(\eta)|^2\right\rangle \cr
&=&2\int d^3\bfk\,P_{\rm(T)}(k)\,\left|T_l^\otimes(k,\eta)\right|^2
\eea
where, $w=k(\eta_0-\eta)$ and $w_0=k\eta_0$, then
\bea\label{LensingTransfer}
&&\hskip-10pt T_l^\otimes(k,\eta)=\sqrt{{1\over2l+1}\sum_{m=-l}^l
\left|\int d^2\hatbfn Y_{(l,m)}^*(\hatbfn)\Omega_{(i)}(\bfk,\hatbfn,\eta)
      \right|^2}\cr
&&=\sqrt{{\pi\over2}{(l+2)!\over(l-2)!}}\,
  \int_0^w dw'\,T(k,{w_0-w'\over k})\,{j_l(w')\over{w'}^2} \ .
\eea
Like $C_l^\otimes$, but unlike $\tilde{\omega}_{(l,m)}$, $T_l^\otimes$ is
invariant to a rotation of  coordinates, and is computed most easily when
$\bfk$ is in the direction of the coordinate ``North Pole''.

	The quantity $\omega$ gives the rotation of the apparent position angle
(PA) wrt the coordinate grid due to the bending of light along the LOS.  This
is related to the apparent shear of the coordinate grid, $\gamma_{ab}$ by the
relation on the sphere
$(\nabla^2+2)\,\omega=-(\gamma^{ab}\epsilon_b{}^c)_{:ac}$.  For this 
$\gamma_{ab}$ to be indicative of the shear measured by looking at galaxy
PA's one requires these PA's not to be preferentially aligned wrt to the
coordinate grid.  However since $\bfH$ at the source (the galaxy) is 
non-zero and non-isotropic, this would not be true if galaxy orientation
were isotropically distributed in physical space.

	Assuming physical isotropy, we must add a ``metric shear'' 
caused by the shearing of the coordinates wrt physical space, {\it i.e.}
$\Delta\gamma_{ab}$, which is just the traceless tranverse projection of
$-{1\over2}\bfH$.  Metric shear does not cause rotation of the images, but to
compare to the above we compute $\Delta\omega$, which is the rotation which
would be caused by the bending of light required to produce
$\Delta\gamma_{ab}$:
\bea
(\nabla^2+2)\,\Delta\omega&=&-(\Delta\gamma^{ab}\epsilon_b{}^c)_{:ac} \cr
&&\hskip-70pt={(\eta_0\cminus\eta)^2\over2}\,
\hatbfn\ccdot\left(\left(\nabla_\perp\cminus{4\,\hatbfn\over\eta_0\cminus\eta}
\right)\ccdot(\nabla\ctimes\bfH)^{\rm T}\right)
\Biggl|_{(\eta,(\eta_0-\eta)\hatbfn)}
\eea
where $\nabla_\perp\equiv\nabla-\hatbfn\,\hatbfn\ccdot\nabla$.  Correcting for
the metric shear leads us to correct the transfer function by adding
\bea
\Delta T_l^\otimes(k,\eta)&=&
{1\over(l+2)(l-1)}\sqrt{{\pi\over2}{(l+2)!\over(l-2)!}}\cr
&&\hskip-50pt
\times\left({l-1\over w}\,j_l(w)-j_{l-1}(w)\right)\,T(k,{w_0-w\over k})
\eea
to eq~(\ref{LensingTransfer}).  Note that none of these metric corrections are
small especially at large redshift and for the longest wavelengths the two
terms nearly cancel.

Physical isotropy need not be the correct approximation as galaxies will 
oscillate in  phase with the gravity waves just as a Weber bar does, and also
because the initial galaxy shapes may have residual correlation with the
gravity wave due to small shear they exert on the galaxy when it is
formed. Just how the galaxy shapes react to these forces depends on the details
of galaxy dynamics and formation.  Here we simply ignore such effects. Both the
metric and induced shear take us away from the ``bending of light'' in flat
space picture for the origin of shear, and while they are negligible in the
case of scalar perturbations, they are not for the very small amplitude of
tensor shear.  

	In the observationally indicated flat FRW cosmology the tensor
metric, $\bfH$, obeys the wave equation
$\ddot{\bfH}-\nabla^2\bfH+2{\dot{a}\over a}\bfH=0$.  With zero
cosmological constant $T_{\rm(T)}(k,\eta)=3j_1(k\eta)/(k\eta)$, and
this is also very for accurate for the observationally indicated
$\Lambda$, given by $\Omega_\Lambda\approx0.7$.

\Sfig{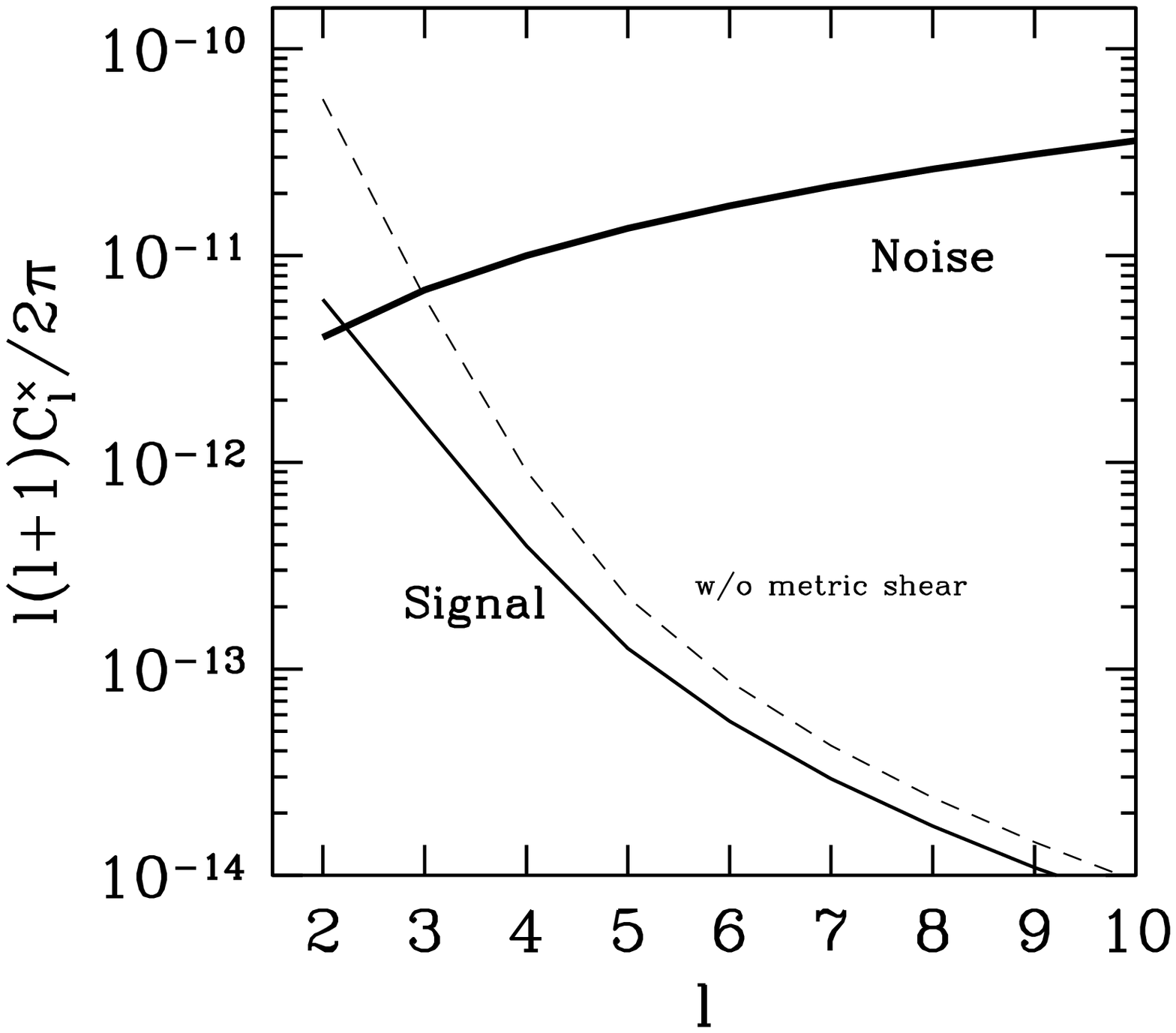}{Expected signal in an all-sky survey in the curl mode in a model
with $\Omega_m=0.3,\Omega_\Lambda=0.7$ with (solid) and without (dashed)
the metric shear term of Eq. (15). The Hubble rate during inflation
which determines the amplitude of the gravity wave power spectrum has been set
to $2\times 10^{14}$ GeV. The noise estimate here assumes an all-sky survey
with $1.5\times 10^{10}$ galaxies and the variance of the 
intrinsic shear equal to
$0.1$. Background galaxies are all assumed to be at fixed redshift  $3$.}

Figure 1 shows the spectrum of the shear produced by inflation. This depends
on the redshift of the background galaxies; here we have chosen a very
optimistic source redshift of $z=3$ for all background
galaxies. The gravity wave shear
$l(l+1)C_l^\otimes$ decreases with increasing $l$, in contrast to density
perturbations which increases with $l$.  The dashed curve in Figure 1 shows
the results using only Eq.~(13), i.e. ignoring the metric shear. We see that
the metric shear correction partially cancels the signal on large scales. So a
very small, virtually undetectable signal becomes completely undetectable
due to the metric shear. Also shown in Figure 1 is the shape noise:
\be
\Delta C_l^\otimes(\eta) =
 \sqrt{2\over (2l+1)f_{\rm sky}} 
 {\langle\gamma^2\rangle\over N_{\rm gal}}
,\eql{noise}
\ee
where $\langle\gamma^2\rangle$ is the intrinsic rms ellipticity of the galaxies 
($N_{\rm gal}$ in all) and $f_{\rm sky}$ is the fraction of sky covered by the
survey.   Figure 1 shows the noise for an all-sky survey with 
$100$ galaxies per square arcminute, or $1.5\times 10^{10}$ galaxies in total,
roughly the density anticipated by the proposed SNAP and LSST missions, though 
they are not all-sky. 

It would be wonderful if inflation-produced
gravity waves produced a C mode of cosmic shear
that could be detected by observing ellipticities of distant
galaxies. Alas, the signal is too small to be detected, and
the best hope of observing C modes remains in the polarization
of the cosmic microwave background.

This work was supported by the DOE at the University of 
Chicago and Fermilab, by NASA grant NAG5-10842 and by
NSF Grant PHY-0079251.

\newcommand\spr[3]{{\it Physics Reports} {\bf #1}, #2 (#3)}
\newcommand\sapj[3]{ {\it Astrophys. J.} {\bf #1}, #2 (#3) }
\newcommand\sprd[3]{ {\it Phys. Rev. D} {\bf #1}, #2 (#3) }
\newcommand\sprl[3]{ {\it Phys. Rev. Letters} {\bf #1}, #2 (#3) }
\newcommand\np[3]{ {\it Nucl.~Phys. B} {\bf #1}, #2 (#3) }
\newcommand\smnras[3]{{\it Monthly Notices of Royal
	Astronomical Society} {\bf #1}, #2 (#3)}
\newcommand\splb[3]{{\it Physics Letters} {\bf B#1}, #2 (#3)}

\end{document}